\documentclass[namedreferences]{solarphysics}
\usepackage[optionalrh]{spr-sola-addons} 
\usepackage{graphicx}        
\usepackage{color}           
\usepackage{url}             
\usepackage[british]{babel}

\begin{document}
\selectlanguage{british}

\begin{article}

\begin{opening}

\title{Variations of magnetic bright point properties with longitude and latitude as observed by \textit{Hinode}/SOT G-band data}

\author{D.~\surname{Utz}$^{1,2}$\sep
        A.~\surname{Hanslmeier}$^{2}$\sep
        A.~\surname{Veronig}$^{2}$\sep
        O.~\surname{K{\"u}hner}$^{2}$\sep
        R.~\surname{Muller}$^{3}$\sep
        J.~\surname{Jur\v{c}\'{a}k}$^{4}$\sep
        B.~\surname{Lemmerer}$^{2}$
       }
\runningauthor{Utz \textit{et al.}} \runningtitle{MBP variations with longitude and
latitude}

   \institute{$^{1}$ Instituto de Astrof\'{i}sica de Andaluc\'{i}a (CSIC), Apdo. de Correos 3004, E--18080 Granada, Spain;
                     email: \url{utz@iaa.es} email: \url{dominik.utz@uni-graz.at}\\
              $^{2}$ IGAM/Institute of Physics, University of Graz, Universit{\"a}tsplatz 5, A--8010 Graz, Austria \\
          $^3$ Laboratoire d\'{}Astrophysique de Toulouse et Tarbes, UMR5572, CNRS et Universit\'{e} Paul Sabatier Toulouse 3, 57 avenue d\'{}Azereix,
             F--65000 Tarbes, France\\
             $^4$ Astronomical Institute, Academy of Sciences of the Czech Republic (v.v.i.), 251 65 Ond\v{r}ejov, Czech Republic
             }

\begin{abstract}
Small-scale magnetic fields can be observed on the Sun in high resolution
G-band filtergrams as magnetic bright points (MBPs). We study \textit{Hinode}/ Solar Optical Telescope (SOT) longitude and latitude scans of the quiet
solar surface taken in the G-band in order to characterise the centre-to-limb
dependence of MBP properties (size and intensity). We find that the MBP's sizes increase and their intensities decrease from the solar centre towards the limb. The size distribution
can be fitted using a log-normal function. The natural logartihm of the mean ($\mu$ parameter) of this
function follows a second-order polynomial and the generalised standard deviation ($\sigma$ parameter) follows a
fourth-order polynomial or equally well (within statistical errors) a sine function. The
brightness decrease of the features is smaller than one would expect from the
normal solar centre-to-limb variation; that is to say, the ratio of a MBP's brightness to the mean intensity of the image increases towards the limb. The centre-to-limb variations of the intensities of the MBPs and the quiet-Sun field can be fitted by a second order
polynomial. The detailed physical process that results in an increase of a MBP's brightness and size from Sun centre to the limb is not yet understood and has to be studied in more detail in the future.
\end{abstract}
\end{opening}

\section{Introduction}
     \label{S-Introduction}
Magnetic Bright Points (MBPs) are thought to be manifestations of magnetic field concentrations
in the photosphere, reaching the kG range. They are small in size (around 200 km in diameter, see \textit{e.g.} \inlinecite{2010ApJ...725L.101A})
and bright (\textit{e.g.} \opencite{2004A&A...422L..63W}). They are best detected on the solar disc
centre (see \textit{e.g.} \opencite{1983SoPh...87..243M}; \opencite{2001ApJ...553..449B}). The contrast of
MBPs compared to the surrounding intergranular lanes is enhanced in the
continuum and is even higher in magnetically sensitive wavelength bands such as the Fraunhofer G-band (see, \textit{e.g.}, \inlinecite{2003ApJ...597L.173S} or
\inlinecite{2002A&A...382..312L} for a detailed study of the dependence of G-band
observations with the heliocentric angle).

Faculae or facular grains (\textit{e.g.} \opencite{1975SoPh...45..105M}) are solar bright features which can be best observed at the solar limb. They are thought to be manifestations of evacuated, mostly vertical flux tubes
(for a theoretical discussion about flux tubes see, \textit{e.g.},
\inlinecite{1984A&A...139..435D}) like MBPs. Due to the strong
inclination of the line of sight to the local normal at the solar limb they seem to be projected
onto neighbouring granules. As the flux tube is evacuated, it appears quasi-transparent and the hot elements lying behind, such as granules, can be seen. This effect is called ``the hot wall effect'' (see \opencite{2004ApJ...607L..59K}).
\inlinecite{2007ApJ...661.1272B} state that faculae are the edge of hot
granules seen through a forest of magnetic elements in plages and network
regions. Nevertheless, these authors also find that
MBPs and faculae are clearly related elements. Earlier studies dealt mostly with the static appearance and properties of faculae (see \textit{e.g.}
\inlinecite{1999SoPh..189...57S}). In recent years the investigation of the dynamics using observations (see \textit{e.g.} \inlinecite{2006ApJ...646.1405D}) as well as MHD
simulations (\opencite{2005A&A...430..691S}) was possible. This could be only achieved due to new instrumentation and methods. The
scientific interest in faculae mainly arises due to their large contribution to
the variations of the total solar irradiance (TSI, see \textit{e.g.} \inlinecite{2004ApJ...611L..57F}
and references therein). The detailed relationship between faculae on
the one hand and MBPs on the other is still not well determined. An interesting recent
study which tries to shed light on these aspects is that of \inlinecite{2009A&A...502..303K}.
These authors developed a method to uniquely
discriminate and classify features on the solar disc as either MBPs or faculae and plan to derive further statistical parameters of MBPs and
faculae, as well as of their interaction in the future.

The photospheric characteristics of MBPs and faculae and their variation with heliographic longitude and
latitude, as well as along the solar cycle are important to obtain deeper insight into the action of the solar dynamo. For example, it is well known
that strong and extended magnetic fields on the Sun (sunspot groups) appear in
so-called activity belts, which tend to shift in latitude, closer to
the equator, during an ongoing solar cycle (see \textit{e.g.} the review by
\inlinecite{2010LRSP....7....1H}). Furthermore, the so-called active longitudes exist; these longitudes feature an enhanced probability for the emergence of new magnetic flux from the solar interior.
Yet, to the best of our knowledge, there are no studies of small-scale solar magnetic fields seen as magnetic bright points (MBPs) in
this context\footnote{There is one interesting study from
\inlinecite{2004ApJ...610L.137C} which deals with centre-to-limb variations on
simulated G-band images.}. This may be due to the lack of time series of
high-resolution filtergrams that scan the solar disc from east to west and from north to south in a systematic way. The situation has recently changed with the advent of new space-borne instrumentation.
We report on an analysis of the variations and dependencies of MBPs with solar longitude and latitude, based on observations
made with the \textit{Hinode}/Solar Optical Telescope (SOT) G-band broadband filter imager (\opencite{2007SoPh..243....3K}, \opencite{2008SoPh..249..167T}). As \textit{Hinode} is orbiting Earth, the
influence of seeing introduced by the Earth's atmosphere is not present and the data sets are stable enough to deduce the
characteristics of MBPs as functions of longitude and latitude. We present results about
the size distribution of MBPs and their mean intensities.
Such information should be useful for theoretical flux-tube modelling.

\section{Data} 
      \label{S-general}
We study two data sets obtained on the 21st and 22nd of December, 2009. The data sets consist of two time series of
high-resolution filtergram data corresponding to an east-west and a south-north scan. Such
scans are routinely taken (on average once a month) and can, in the future, provide valuable information about changes along the solar-cycle. All filtergrams are taken in the G-band with the broadband filter imager (BFI) of
\textit{Hinode}/SOT. The filter is centred around 430.5 nm (CH I molecular lines;
magnetically sensitive) and has a full width at half maximum of 0.8 nm.
\subsection{East-West scan} 
  \label{S-text}
The east-west scan was taken on the 22nd of December, 2009.
All images are binned in 2x2 pixels and have a field of view (FOV) of about 220 arcsec
per 110 arcsec. The spatial sampling is 0.108 arcsec/pixel in both, the $x$ and $y$, directions. The complete scan consists of 20 images. Each of the 10 different positions
along the solar equator was imaged twice. The scan started on the 22nd of December at 10:22 UT at an eastward position of -920 arcsec (compared to the
disc centre) and moved step by step (nominal step-size of 200 arcsec)\footnote{Around
the central position the step-size was about 150 arcsec (from -123 arcsec to 76 arcsec).}
on to the furthest westward longitude of
870 arcsec, which was reached at 14:22 UT. The time between the two positions was roughly 25 minutes and it took about 3 minutes to take the two exposures.

\subsection{South-North scan}
The programme
started at 14:30 UT on the 21st of December at -930 arcsec southward of the centre of the
Sun and finished at 2:13 UT of the 22nd at a position 930
arcsec northward. As in the former scan, all images are binned in 2x2 pixels
and have a field of view (FOV) of about 220 arcsec per 110 arcsec. The temporal
coverage corresponds to two latitudinal positions per hour (nominal step-size of about 100 arcsec). For every latitudinal position the
scanning programme takes 2 images with a temporal difference of about 3
minutes between the exposures. As the FOV of the instrument is rectangular
with an aspect ratio of 2 to 1, a complete scan from south to north consists of
twice as many images as an east-west scan, thus amounting to 40 exposures.
The time lag between south-north and east-west scans is as small as possible in order
to settle a reasonable basis for future comparative and long-term studies of these quantities.

\section{Analysis}
Both data sets were analysed in the following way. First of all the primary
reduction and calibration (including flat-fielding, dark-current subtraction,
correction for bad pixels and read-out errors) were done through the corresponding
SSW\footnote{SSW stands for Solar Software. It is an add-on software
package for the programming language IDL specially dedicated to solar
physics. The complete description of the \textit{Hinode}/SOT analysis guide can be found at: \url{http://hinode.nao.ac.jp/hsc_e/analysis_e.shtml}.}
IDL routines. The identification of the MBP features was done with a fully
automated algorithm presented in \inlinecite{2009A&A...498..289U}. The algorithm segments
the images and subsequently identifies the interesting features. The
identification uses one of the MBP key features, namely the steep local
brightness gradient. The algorithm was developed originally for the identification of MBP features and exhaustively tested on data taken at solar disc centre. Nevertheless, it might happen that facular features close to the solar limb are partly identified too.

The feature size is then calculated by applying a full-width at half-maximum
(FWHM) criterion on the candidate MBP segment. In our case the median intensity of the pixels belonging to the boundary of a segment is calculated. All
pixels exceeding the brightness of the median boundary intensity plus half of
the difference to the maximum intensity are considered for calculating the size
of the feature. The position of the features on the solar disc
has to be converted to a heliocentric angle. This angle enables a
correction of geometrical projection effects (increase of pixel size towards the solar limb).

From both data sets we derived the following MBP quantities:
\begin{enumerate}
 \item Size distribution of MBPs.
 \item Log-normal fit coefficients of the size distribution.
 \item Mean intensity of the MBP features, as well as of the full image.
 \item Intensity ratio between the MBP features and the mean image
intensity. 
\end{enumerate}

The positions of the features on the solar disc were converted to heliocentric angles. Their sizes were then compensated for projection effects with the help of those angles and taken as the diameters of circles with an equivalent area\footnote{It is therefore implicitly assumed that MBPs are roundish
features. Another possibility would be to estimate the north-south (for the east-west scan) and east-west (for the north-south scan) elongation
of the features. These feature lengths are not influenced by the foreshortening effect and provide a direct measurement of the size/extension of the features. Unfortunately, this method is not applicable to MBPs due to their small size and, hence, the low number of available measurement points (about 3 different possibilities for a spatial sampling of 0.1 arcsec).}. The diameters of the features can be plotted and interpreted
subsequently. We fitted the size distribution with log-normal functions:
\begin{equation}\label{log_normal}
    f(\mu,\sigma,\omega)=\frac{\omega}{x \sigma \sqrt{2 \pi}}\exp\left(-\frac{(log (x) - log(\mu))^2}{2 \sigma^2}\right),
\end{equation}
where $\omega$ is the normalisation or weighting parameter, log $\mu$ is the mean value of the natural logarithm of the data values (sometimes called location parameter) and $\sigma$ is the generalised standard deviation (sometimes refered as scale parameter) known from the normal
distribution. Earlier works, such as
\inlinecite{2010ApJ...722L.188C} and \inlinecite{Otmar}, already recognised the possibility to fit the size of MBPs by log normal distributions. The mathematical concept behind a log-normal distribution is that
the logarithm of the $x$ values follows a normal distribution (instead of the
$x$ values themselves). From the physical point of view the normal distributions
are generated by additive processes and, hence, yield an additive error and
variance, whereas log-normal distributions appear in or are typical of physical processes
based on multiplicative laws (\textit{e.g.} those found in growing processes) and, thus, give a
multiplicative variance. We expect a log-normal distribution for
the size of MBPs as they regularly undergo fragmentation and merging processes
(see \inlinecite{2010ApJ...722L.188C}; or \inlinecite{1988ApJ...327..451B} for an analytic interpretation of
fragmentation processes and their relation to the log-normal distributions). More details and background information about
the log-normal distribution can be found in \inlinecite{Eckhard}.

For further analysis of the log $\mu$ and $\sigma$ parameters, we only considered those values
with a reduced $\chi^2$ fit quality within a range of 1.0 $\pm$ one standard deviation of
all reduced $\chi^2$ values obtained by the size distribution fits. In the following,
polynomial fits based on these parameters were obtained. In order to improve the
quality of these fits and to consider the
$x$-axis error, we used the following recursive scheme:

The $\Delta x$ (uncertainties of x) were propagated using the obtained polynomial functions to
transfer them into a corresponding $\Delta y$ error: 
\begin{equation}\label{error_prop}
    \Delta y_{p}=f'(x) \cdot \Delta x_m,
\end{equation}
where $\Delta y_{p}$ is the propagated $y$ error due to $\Delta x_m$, which is the measurement uncertainty in x and $f'$ is the derivative of the fitting function (in our case a polynomial).
The original and the
propagated errors ($\Delta y_m$, $\Delta y_p$) were squared and summed quadratically to get an
effective $y$ error ($\Delta y_e$). Thus, considering now errors in $y$ as well as in $x$: 
\begin{equation}\label{error_eff}
    \Delta y_{e}=\sqrt{{\Delta y_{p}}^2 + {\Delta y_m}^2}.
\end{equation}
These errors were used to fit again the log $\mu$ as well as the
$\sigma$ parameter and, hence, improve the quality of the $\chi^2$ goodness. If the obtained value for the $\chi^2$ goodness is still not acceptable, the process can be repeated until a
stable and acceptable $\chi^2$ value is reached (for detailed information see
\opencite{1982AmJPh..50..912O}).

The mean image intensity and the contrast of the images
were calculated by excluding dark background pixels beyond the solar
limb. This was done using a simple threshold method. All pixels below a value of
100 data units in intensity were considered as dark background pixels and precluded from the analysis.

The MBP intensity for a given position was calculated by taking the average of the most intense pixel
of the MBP features in a given image. The centre to limb intensity distribution for the images (mean image intensity), as well as for the
MBP features was then fitted by a second-order polynomial. In a next step the
intensity ratio between these features and the mean image intensity was
calculated by dividing the two fits for the centre to limb variation.


\subsection{Results for the East-West Scan:}

\begin{figure*}
    \centering
        \includegraphics[width=0.8\textwidth]{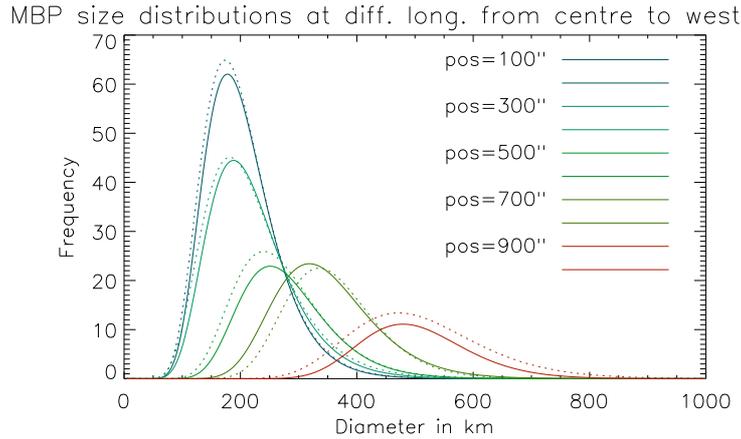}\\
        \caption{The measured size distribution of MBPs for different longitude positions on the solar disc.
The distributions follow log-normal functions and were obtained to the west of the
solar disc centre. Positions as are indicated by the legend (values are in arcsec). The
reduced $\chi^2$ value of the fits is $1.16\pm 0.35$. At each position the size distribution was obtained twice (coloured pair in full line and dashed line, respecitvely) by using the twin images.}
    \label{size_dist}
\end{figure*}

\begin{figure*}
    \centering
        \includegraphics[width=0.99\textwidth]{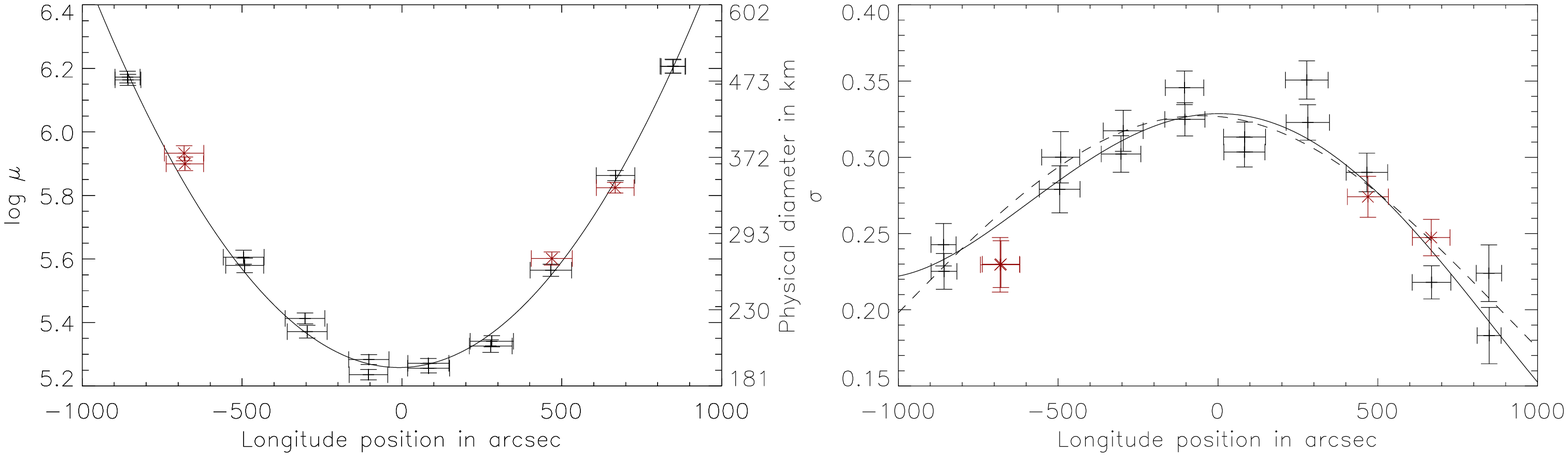}
        \caption{Left: The log $\mu$ parameter of the log-normal distributions shown in Figure \ref{size_dist} displayed versus longitude together with a second-order polynomial fit (solid line). Points marked in red have a $\chi^2$ value
outside one standard deviation interval around the optimal value of 1 and are not considered to obtain the plotted curve. The reduced $\chi^2$ of
the fit is 0.4. Right: The parameter $\sigma$ of the size
distributions shown in Figure \ref{size_dist} versus the longitude together with a fourth order polynomial (solid line) and a sine function fit (dashed line). Values with a low confidence value
($\chi^2$ outside the acceptable range) are again shown in red
and not used for the fit. The reduced $\chi^2$ value for the polynomial is 2.4, while for the sine function it takes a slightly better value of 2.3.}
    \label{fit_para_size}
\end{figure*}

The size distribution of MBPs during the east-west scan displays an increase in
size and a decrease in frequency (number of detected features\footnote{Keep in mind that we have not normalised the number of detected
MBPs to the size of the FOV, i.e. the decrease of detected features towards the limbs per unit solar surface is even more pronounced.}) when approaching
the limb (see Figure \ref{size_dist}). All the size distributions were fitted by log-normal curves. Figure \ref{fit_para_size} shows the parameters of the size distributions. The log $\mu$ parameter was fitted by a second order polynomial, the reduced $\chi^2$ value was 0.4. The $\sigma$ parameter was fitted by a fourth order polynomial (solid line) and a sine function (dashed line); the reduced  $\chi^2$ values were 2.4 and 2.3 for the polynomial and sine function, respectively. The position values for the parameters are the mean MBP $x$-positions within the FOV. Hence, the error bars in
the plots represent the standard deviation of these mean positions. Due to the
fact that the number of detected features as well as their spatial distribution
on the disc changes from image to image, these error bars are not constant and
the average $x$-position is not necessarily coincident with those of the
images. The good agreement (same results within errors) between independent measurements in the two exposures taken at the same solar disc position is a good consistency test for the quality of the data sets and the analysis pipeline. Our plots show only one half of the results as the curves for the second half resemble the first and would not provide additional
information but make the plots appear overloaded (4 similiar curves per measurement position). To keep the plots clearer, we also omit the original data points in Figure \ref{size_dist} and show only the respective fits.

\begin{figure*}
    \centering
        \includegraphics[width=0.99\textwidth]{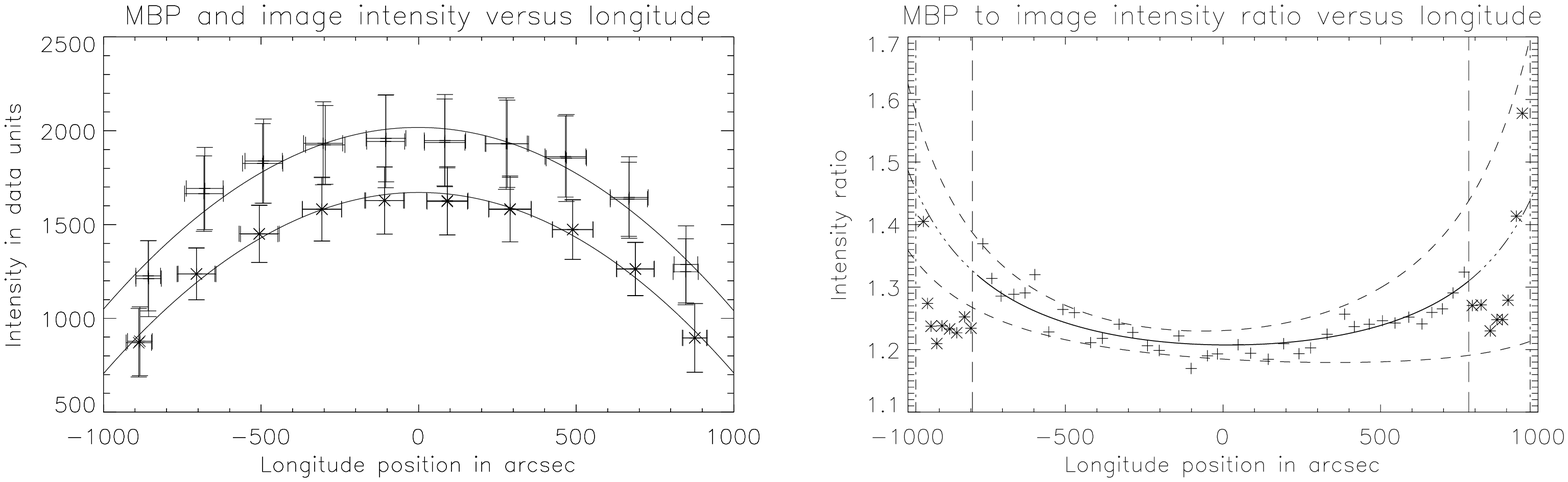}
        \caption{Left plot: The mean image intensity (lower
curve) shown together with the mean MBP intensity upper curve versus longitude. The fit applied to the measurements is a second order polynomial.
Right plot: The ratio between the
mean MBP intensity at a given longitude to the mean image intensity at the
same longitude. The curve represents the ratio of the fits shown to the left. The dashed
lines show the 1-$\sigma$ interval around the resulting curve. The
crosses give values obtained by local calculation of the ratios (\textit{i.e.} by
dividing the MBP intensity by the median intensity of the surrounding). Good
MBP statistics is found for points inside the interval marked by the long-dashed vertical lines. Measurements (asterisks) outside this interval have to
be treated with care. The
actual size of the solar disc is marked by the dash dotted lines around 980
arcsec.}
    \label{intens}
\end{figure*}

The mean intensity of the image, as well as of the MBP features, decreases towards the
limb and can be fitted by a second order polynomial (see Figure
\ref{intens}, left panel). One has to bear in mind that the MBP positions, as given in the plot, are mean values obtained by averaging over all the positions of the detected MBPs in the FOV. Therefore, these points can show a slight deviation when compared to the positions in corresponding images. Other studies, concentrating more on the limb darkening aspect,
have used more sophisticated models and hence higher-degree
polynomials (\textit{e.g.} fifth order polynomials; \opencite{1994SoPh..153...91N}). From the right panel of
Figure \ref{intens}, it is obvious that the intensity of the MBP features
decreases more slowly than the mean image intensity when approaching the
limb, which gives rise to an increasing ratio. We also obtained the
ratio via a direct comparison of the MBP intensity to the local
surrounding intensity (median intensity within a 20 arcsec per 20 arcsec subfield) shown with crosses. Up to 800 arcsec from the solar disc centre both methods (the centre to limb variation (CLV) intensity ratio and local ratio method) agree within errors (crosses lie within dashed lines). At around this distance from disc centre the
increase to the limb ceases (interval indicated by long-dashed lines). This is
most probably due to identification problems or to the disappearance of MBP
features when approaching the solar limb\footnote{\inlinecite{2009A&A...502..303K} state that MBPs are concentrated more towards solar disc centre whereas faculae tend to appear closer to the solar limb.
The details of the spatial distribution of both features and their intersection are
yet to be investigated in detail.}. The number of detected MBPs decreases
steadily and drops below 66\% of the mean number of identified features per
image at distances of 800 arcsec from disc centre and beyond. At this distance a significant change can also be seen in the size
distribution and its behaviour. Therefore, local ratio points outside this range should be treated carefully and are
marked as asterisks. The curve is asymmetric, as found for
the contrast as well, and these phenomena may be related (see Section 4).

\subsection{Results for the South-North Scan:}

\begin{figure*}
    \centering
        \includegraphics[width=0.8\textwidth]{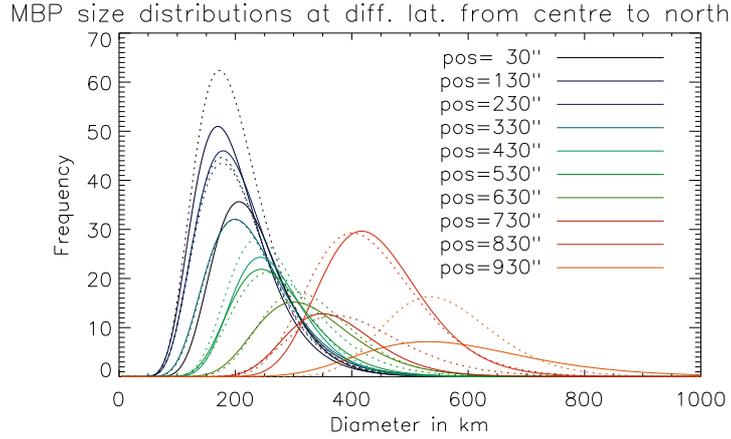}\\
        \caption{The size distributions obtained
during the south-north scan. A continuous shift to larger sizes with increasing
distance to the centre can be seen. The reduced $\chi^2$ value is cloes to 1 with a one $\sigma$ interval of $\pm$ 0.37. The legend gives
the position at which the distributions were obtained in arcsec measured from
the disc centre northwards. For each position two independent images were analysed (shown in same colour in dashed and solid line).
}
    \label{size_dist_ns}
\end{figure*}

\begin{figure*}
    \centering
        \includegraphics[width=0.99\textwidth]{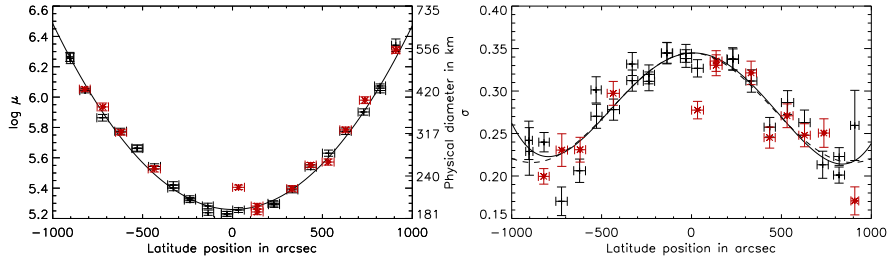}
        \caption{Left: The log $\mu$ paramter of the log-normal distributions shown in Figure \ref{size_dist_ns} is displayed versus the
latitude. The log $\mu$ parameter versus latitude was fitted using a second order
polynomial. Values in red correspond to a reduced $\chi^2$ outside the one $\sigma$ interval around 1. Right: $\sigma$ parameter of the distributions shown in Figure \ref{size_dist_ns} versus latitude. In this case the parameter was fitted using a fourth order polynomial (shown in solid line) and a sine function (shown in dashed
line). Values in red have the same meaning as before.}
    \label{ns_fit_para_size}
\end{figure*}

The size distributions obtained from the south-north scan (see Figure \ref{size_dist_ns}) display a similar behaviour as those obtained from the east-west scan. Namely, an increase in size and a decrease in number of features as one approaches the poles.
In particular, the number of detected features with smaller sizes is decreasing while the number of larger features stays nearly constant.
This is probably an indication of a selection effect (for more
details, see Section 4). For the centre-to-north distribution the change in the fitting parameters is not as steady and monotonical as for the centre-to-west
distribution. Specifically, the 2 distributions derived at 830 arcsec, far from disc centre, are outliers. Furthermore, one of the 930 arcsec distributions does not
follow the steady trend. This might be just due to the weak MBP statistics (low number of identified features) close to the poles. The log-normal parameters of the size
distributions are shown in Figure \ref{ns_fit_para_size}. The left panel
illustrates a trend of increasing sizes towards the poles given by the log
$\mu$ value. Furthermore, the curve is symmetric and can be well fitted by a second order
polynomial. After considering and propagating the error in
position (as described before), the reduced
$\chi^2$ has a value of 0.89. 
The parameter
$\sigma$ was fitted using a fourth order polynomial. The fit follows
the measured values as well as in the case of the east-west scan with a reduced
$\chi^2$ value of 2.25, after consideration of the position errors. 
The second fit shown with a dashed line belongs to a sine function with a
corresponding goodness about 2.2.

The image as well as the MBP intensity behaves similarly to the
intensities found in the east-west scan. The polar images and features are
clearly darker than the ones at the centre of the solar disc (see left plot of Figure \ref{intensity_ns}). The ratios between the
mean feature intensity to the mean image intensity are also very similar. In the case of the
east-west scan the ratio reaches a value of 1.45, while in the case of
the south-north scan it goes up to a value of about 1.55 (see right plot of Figure
\ref{intens} compared to the right plot of Figure \ref{intensity_ns}). This can be due to
averaging over different FOVs (the used FOV was rectangular). To
verify the results found computing the ratio between the intensity curves we used a second method as described before (shown by crosses and asterisks; asterisks represent the results of the second method with weaker statistical significance). As the Sun was very quiet at the time of
observations there was no need to exclude active regions. The two independent methods yield similar results agreeing within the errors. Furthermore, also in this case the CLV of the intensity ratios shows an asymmetry with
higher intensities in the first measured hemisphere (southern) and larger errors in the boundaries in the second hemisphere (northern).

\section{Discussion}
\subsection{Data sets}
The data sets were acquired during a period of very low magnetic activity. Therefore, the spatial distribution and the properties of MBPs should not be influenced by active groups and regions. The solar rotation could affect the analysis by transporting the same features several times in the observed FOV but this process can be neglected as the lifetime of MBP features is in the range of several minutes (see \textit{e.g.} \inlinecite{2005AA...441.1183D}) while the solar rotation takes place in the range of days. Furthermore the scanning of the solar surface takes from east to west and south to north only a few hours which is too less time for the solar rotation to cause significant changes. In addition the solar rotation and the so-called limb effect (see \inlinecite{2002A&A...382..312L}, their Figure 10) could shift the CH molecule lines slightly in the wavelength domain.
The broad width of the used G-band filter (0.8 nm) precludes any significant effect coming from those line shifts.
For the positioning and accuracy of the satellite pointing we had to
rely on the operators. This means that all positioning values have an
additional error coming from the spacecraft positioning precision. In the
instrumental paper by \inlinecite{2007SoPh..243....3K} (page 10, Table 3) the
values for the absolute pointing are specified as 20 arcsec and the pointing
determination as 0.1 arcsec. Also the drift of the satellite within a taken
position is negligible (0.6 arcsec/2 seconds).

We did not incorporate images corrected for the centre-to-limb variation of the intensity in our analysis as we wanted to apply the algorithm on images as
close as possible to the original ones. Additionally, different procedures,
methods and centre-to-limb variation fitting functions exist in literature (see
\textit{e.g.} \inlinecite{2007ApJ...661.1272B}, \inlinecite{2005A&A...438.1059H} or \inlinecite{1994SoPh..153...91N}).
Therefore, different outcomes may be caused by the incorporation of different
data reduction steps. One has to have in mind that every step of data reduction
(dark current, flat fielding, bad pixel correction and so on) always bears the
danger of artificially modifying the data and incorporating additional errors.

\subsection{Algorithm}
Since the used identification algorithm works with local brightness gradients the
achieved image contrast is of high importance. Therefore, it is
necessary to have good and especially stable image contrasts during the
complete scans. This can be seen in Figure \ref{scatter_nr} that illustrates the
ratio of image contrasts versus the number of detected MBP features in subsequent
exposures. Here, we use the advantage of the data sets that have a double
exposure per measured solar disc position. Knowing that the exposures were
taken at the same position and nearly co-temporally (3 minutes time
lag) a possible change in image contrast can only be due to slight variations
of the solar features and more likely due to instrumental effects. Changes in the
number of detected features are therefore related to changes in image quality.
This helps us to estimate the influence of the image contrast on the
detection quality. The figure shows the expected correlation between increasing
contrasts and number of detected features (correlation coefficient of about
0.4). A perfect algorithm should be independent of image contrast variations
and, hence, the scatter should be equally distributed in all 4 quadrants and the
correlation coefficient should be very close to 0. We tested in the same way a
possible correlation between MBP intensity and contrast, as well as between MBP area and
contrast. Both scatter plots (also shown in Figure \ref{scatter_nr}) yield no distinct
relationship (the values are distributed more or less evenly over all 4
quadrants). This plot was done for the south-north data set due to the better statistics and higher accuracy (twice more observational positions compared to the east-west scan). A similar behaviour can be expected for the east-west scan.

\subsection{Image stability}
Due to the importance of image contrasts on the stability of the algorithm we investigated
them in more detail. The image contrast is given by the standard deviation of
image intensity divided by the mean intensity. Figure
\ref{data_1} displays the image contrast of the east-west scan. A
tendency for higher contrasts from the west side compared to
those from the east side can be identified. This can have a physical/solar origin (more active groups or magnetic activity on one side) or can be due to instrumental reasons. As the Sun was very quiet at the observational
time we rather tend to support the instrumental origin  having in mind the existence of changes due to thermal stress during the scan. The spacecraft expands and shrinks according to
the heat load; this leads to a change in the optimal focus position (hence
affecting the image contrast) from the centre towards the limb. Insets in Figures \ref{data_1} and \ref{data_2} highlight the variations of the image contrast for scans close to the disc centre.
\begin{figure*}
    \centering
        \includegraphics[width=0.99\textwidth]{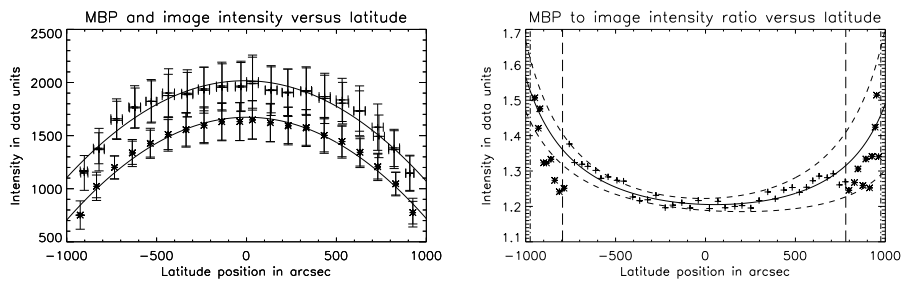}
        \caption{Left plot: The mean image intensity (lower curve) and the mean MBP intensity (upper curve)
versus the latitude is displayed for the south-north scan. See also  Figure
\ref{intens} left plot. Right plot: The intensity ratio between the MBP intensity and the
mean FOV intensity is shown. For more information see Figure \ref{intens} right plot and the description in the caption.
}
    \label{intensity_ns}
\end{figure*}


Figure \ref{data_2} illustrates the contrast of the images along the
south-north direction. The measurements follow a parabolic curve and are
much more symmetrically distributed than in the case of the east-west scan. Therefore, we can conclude for this data set that the thermal
equilibrium and heat load relaxation was nearly achieved (there was nearly no contrast asymmetry due to instrumental effects between the southern and the northern hemisphere).

Such contrast asymmetries may explain the asymmetry in the CLV of the intensity ratio (see Figsures \ref{intens} and \ref{intensity_ns}). A smaller contrast in an image may be explained by de-focusing (smearing) which leads to smaller intensity differences and also to smaller absolute intensities when compared with a focused image. In both figures an increase of the contrast to the limb can be seen. We did not correct the data for quiet Sun CLV of intensities to avoid the introduction of additional errors. This causes, on the other hand, the observed rise of contrast close to the solar limb.
\begin{figure*}
    \centering
        \includegraphics[width=0.8\textwidth]{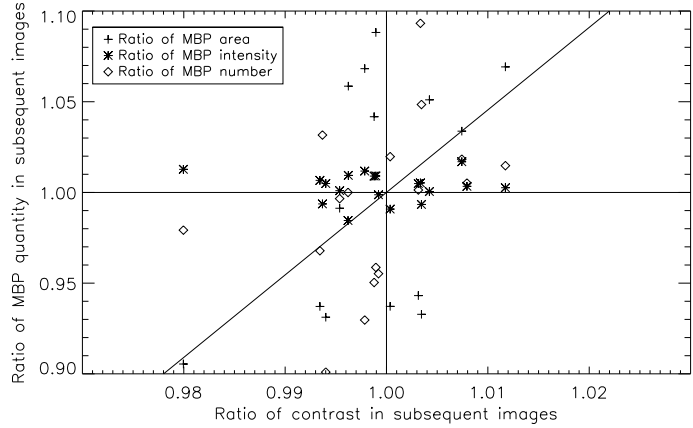}\\
        \caption{Scatter plot of the ratio of image contrasts to the ratio of the number of detected MBPs in subsequent exposures (diamond symbols) for the south-north data set. The solid line gives a least square linear fit passing through the (1,1) point. The slope of the fit
illustrates the influence of the image contrast on the detection probability (1\% change in image contrast results in about 4.5\% change in detected MBPs). A
horizontal line would represent a perfect algorithm whose detection probability
is independent of the image contrast. Furthermore, the scatter plots of the ratio of image contrasts to the ratio of detected mean MBP intensities (asterisks) and mean areas (crosses) in subsequent exposures are shown.}
    \label{scatter_nr}
\end{figure*}


\begin{figure*}
    \centering
        \includegraphics[width=0.8\textwidth]{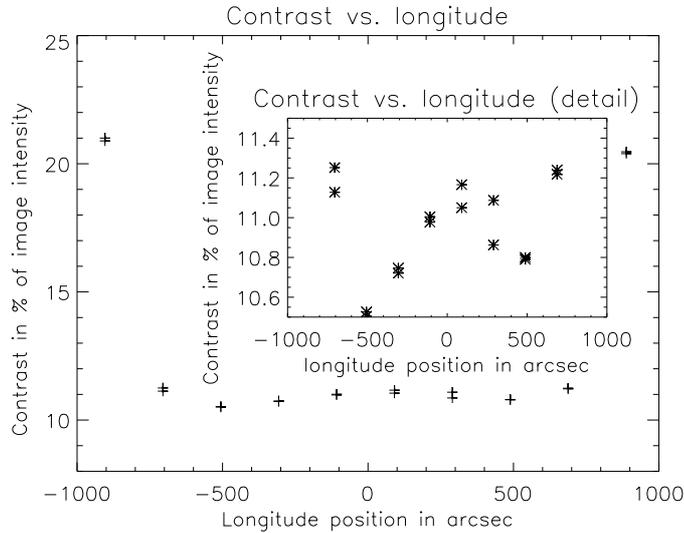}\\
        \caption{The image contrast during the east-west scan I. To highlight the
smaller contrast changes in the centre of the disc, a zoom is shown as inset. A tendency towards higher contrasts on the west side compared to the east side can be identified.}
    \label{data_1}
\end{figure*}

\begin{figure*}
    \centering
        \includegraphics[width=0.8\textwidth]{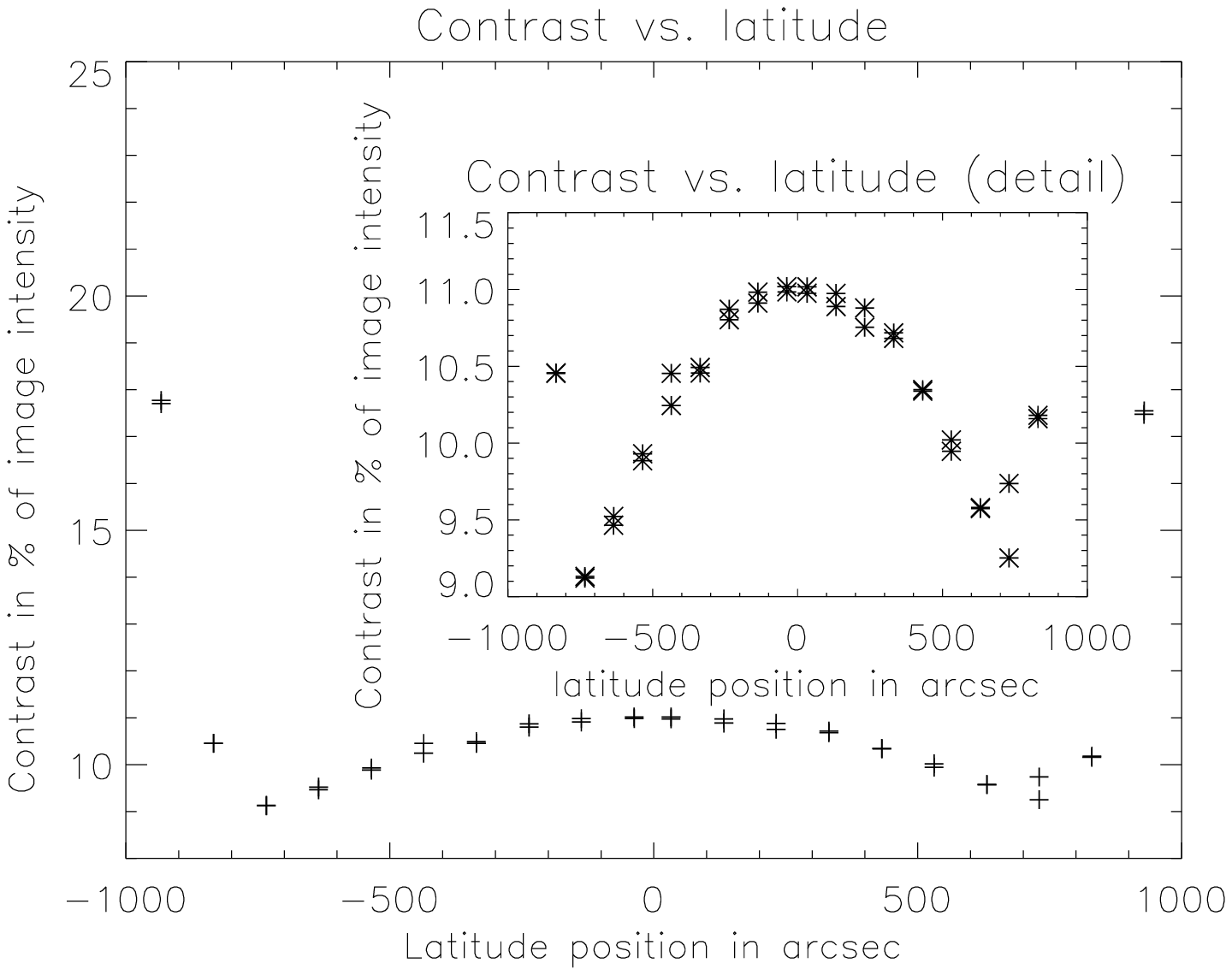}\\
        \caption{The image contrast during the south-north scan II is shown. The smaller contrast changes in the middle of the disc are shown with a zoom and displayed as inset. The contrast is clearly higher during the second part of the
scan (northern part).}
    \label{data_2}
\end{figure*}



In summary the instrumental contrast variations should stay below 10\%. Considering a variation of 5\% in the number of detected features per 1\% variation
of image contrasts (values taken from Figure \ref{scatter_nr}), the total variation of detected elements due to the
interplay between the algorithm and instrumental effects can result in a quite huge
factor of nearly 50\% reduction in the detection probability at the limbs. Yet the impact on
the interesting parameters for the present study (intensity, size) is well
below that value as the algorithm does not show any preferential selection
effects with the tested varying image contrasts\footnote{Only small variations
have been tested due to the subsequent image condition. On larger scales there
might be a measurable influence of image contrasts on the intensity/size of the
detected features.}. For future works regarding the variation of the number of MBPs
the interplay between image contrasts and detection probability should be considered.

\subsection{Field of view effects}
Images from the analysed data sets
have the same size, but due to their rectangular shape the obtained measurements are averaged over a different
 range of longitudes and latitudes. This may lead
to different results even though the physical appearance of the observed solar
surface is the same. We investigated the behaviour of the east-west scan G-band
images varying the sizes of the FOVs (\textit{i.e.} splitting the complete FOV in
several subfields).

We find a major change occuring in the results between the original FOV compared to a half sized
FOV. Later on, the changes are only minor and might not affect the outcome of the presented and/or similar studies.
Therefore, it is important to note that one should work (at least
close to the limb) with quantities averaged over FOVs smaller than 100 arcsec
in length. Larger averaging windows might have a strong influence on the
outcome.
\begin{figure*}
    \centering
\includegraphics[width=0.8\textwidth]{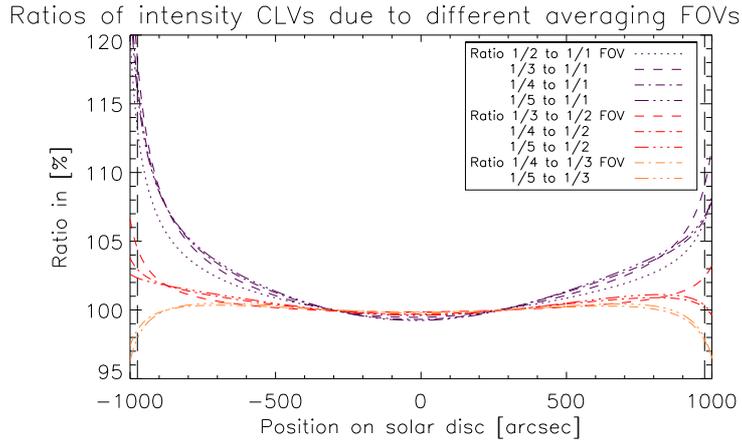}\\
        \caption{The ratios between the intensity CLVs for different sized averaging windows are given. Lines displayed in purple indicate the ratio in \% between
the usage of the full FOV for averaging compared to smaller FOVs. The ratios
are coded as follows: short dashed line, ratio to an averaging window over 1/2 of the
FOV; long dashed line, ratio to an averaging window over 1/3 of the FOV; long-short
dashed line, ratio to an averaging window over 1/4 of the FOV; long-tripple-short dashed
line, ratio to an averaging window over 1/5 of the FOV; The FOV has a size of 220 arcsec
per 110 arcsec. The color code represents: purple: ratios between different averaging windows to an averaging window over
full FOV; red: ratios between different CLVs to a CLV gained by averaging over half FOV; orange: ratios compared to averaging over 1/3 of the FOV; The dashed vertical lines give the solar limb
at the day of observation.}
    \label{intensity_clv1}
\end{figure*}
This can be seen in Figure \ref{intensity_clv1} in more detail. Here, we
investigated the ratios between the CLVs of the intensity for different
averaging windows. For that purpose we fitted fourth order polynomials to the CLVs of the intensity obtained for averaging windows of different sizes. In the next step we derived the ratio of those polynomial fits. Apparently a rather large
deviation between the obtained CLVs occurs especially at the eastern limb (when
averaging over the full FOV, \textit{i.e.}, 220 arcsec per 110 arcsec) compared to the next
smaller averaging window of 110 arcsec per 110 arcsec (all lines coloured in
purple represent comparisons between full FOV and these smaller FOVs). A further
decrease of the size of the window does not change the results significantly (see, \textit{e.g.}, red and orange lines; differences are smaller than 5\% even at the limb). Therefore, we suppose that possible differences between the
brightness enhancement of MBPs (ratio of MBP intensity to local image intensity) for the south-north scan (polar) compared to the east-west scan (see right plots in Figures 
\ref{intens} and \ref{intensity_ns}) are mostly due to this effect.

\section{Comparison with previous works}
\subsection{Size distributions}
We have seen an increase of the size of MBPs when approaching the limb. This
increase can have several origins. The most important and
apparent one is the effect of geometric projection. This means that when looking at the flux tube from a large angular direction, the visible portion of the hot wall increases and the complete structure appears larger and
brighter. This effect might also explain the increase in brightness relative to
the surrounding for the identified structures closer to the limb (see papers
about the hot wall effect, \textit{e.g.}, \inlinecite{2004ApJ...607L..59K}). Another
interesting possibility to explain this effect would be the characteristics of
the flux tube itself. As it is a partially evacuated structure, it is
possible that the detected features correspond rather to parts of granules which
appear brighter through the transparent magnetic elements than the magnetic feature itself (see e.g
\inlinecite{2005A&A...438.1059H} and references therein). Further sources
influencing the detection and analysis of MBPs are the instrument, as
well as the identification algorithm. In the study of
\inlinecite{2009A&A...495..621C} it was shown that the contrast of
agglomerations of magnetic flux tubes varies with the heliocentric angle and additionally with
the number of flux tubes contained in an agglomeration. When approaching the solar
limb the contrast of a single flux tube decreases faster than the contrast of a
group of flux tubes. Due to the fact that the identification algorithm is
based on a local brightness criterion, \textit{i.e.}, larger features (agglomerations of
not resolved flux tubes) are preferentially detected or \textit{vice versa} smaller
elements (single flux tubes) lack the necessary contrast to be identified.
Furthermore, the contrast of the images varies during the scan with a slight
trend to have a better focus during the end of the scan.
Additionally, every identification (automated or manually) is based on certain
criteria. In our case the identification works on local brightness gradients
(see \opencite{2010A&A...511A..39U}) that result in the change of the identification probability due to the variations in contrast. Finally, one should have in
mind that due to geometrical projection effects the sampled size of a pixel is
changing with latitude and longitude. This can also lead to a preferential
identification of larger features and/or to a shift of the size distribution as
discussed in \inlinecite{2009CEAB...33...29U}.

\subsection{CLVs of the intensity}
We have seen in section 4.4 that the CLVs of the intensity depend on the size of the averaging window. Nevertheless
the presented results are reliable in the sense that deviations only
occur to a larger extent at the limbs and the difference of about
10\% is still acceptable when compared to other influences and to different
results in literature. In the work of \inlinecite{2005A&A...438.1059H} a rather
flat intensity CLV is reported for facular grains (see Figure 5 of the quoted
work), while \inlinecite{1991A&A...246..264A}, as well as
\inlinecite{1978PASJ...30..337H}, reported a slight decrease of the CLV of the intensity of network bright points (a synonym used for MBPs in earlier works) towards the solar limbs. This
difference of the behaviour of the CLV at the limb in the previous studies might be due to different methods (visual identification compared to automated one)
and fitting routines (ratio of second order polynomials to a third order polynomial
fit).

Furthermore, a fourth order polynomial applied to the CLV of the image intensity (used in Section 4.4 and Figure \ref{intensity_clv1} for the quiet Sun)
agrees clearly in a smoother way with the data points than a second order
polynomial does (as used for the MBP CLV and also for Figures \ref{intens} and \ref{intensity_ns}). On the other hand, a higher
degree polynomial features a lower statistical confidence when compared to a simpler polynomial.
This is due to the smaller amount of residual degrees of freedom (namely the number of data points - 1 - degree of the polynomial). In our case the data consists of 10
positions for the east-west scan and 20 positions for the north-south scan which constrains us to a low degree polynomial. It would be possible to split
the FOV in several subfields to increase the amount of data points.
Unfortunately, this would only reduce the number of detected features in the
subfields (especially close to the limbs, where the number is small anyway) and, therefore, the gain in positional accuracy would be lost in the frequency (number of detected features) and, hence, intensity
accuracy. A future step to improve the CLV of MBP
intensities would be the incorporation of several scans and, hence, more data. The problem that will arise is the incorporation of images with
different quality obtained by the telescope at different observational dates.
On the other hand, statistically such different effects may cancel each other.

Finally, we want to remark that for distances of up to 700 arcsec off the solar
disc centre the results for the CLVs of the intensity of MBPs are
generally in good agreement with previous studies such as those of \inlinecite{2005A&A...438.1059H}, \inlinecite{1991A&A...246..264A} and \inlinecite{1978PASJ...30..337H}. Further limbwards, the results differ from study to study most
probably due to the methods, data sets at hand, and the low number of accessible
features.

\section{Conclusions and Outlook}
In this study we have investigated the dependence of MBP characteristics such as
size and intensity with longitude and latitude for quiet Sun
data sets. For the identification of the features, we used a fully automated
detection algorithm on the best available data sets coming from a stable,
space-borne telescope. We found that the detected features increase in size when
approaching the solar limb. This effect was found for the east-west direction, as well as for a south-north scan. The size distribution can be well fitted
by a log-normal distribution. The behaviour of the two parameters of this
distribution can be described by polynomials and for the $\sigma$ parameter
by a sine function equally well. The intensity of the detected features decreases to
the solar limb but slower than the mean image intensity, so that the brightness
enhancement (ratio of MBP intensity to local quiet Sun intensity) increases. This might be due to the inclination of the flux tubes which causes a larger proportion of the hot walls to be visible. At the limb the results differ
from former studies reporting flat intensity CLVs or slightly decreasing CLVs.
The difference might not have a physical origin but rather be due to the different
methods employed in the various studies. Another important point is the effect the size of the averaging window
can have on average quantities. One has to consider this effect, especially
when using FOVs of more than 100 arcsec in length close to the solar limb. Furthermore, the significance of the
results at the limbs is not as high as for values obtained within 800
arcsec from disc centre due to the lower number of identified features. Due to these reasons, and the afore mentioned different outcomes in previous studies, a final conclusion of the behaviour at the extreme limb is therefore outstanding and
a visual verification of the detected features at those positions may help to clarify the picture. The effects found for the size distribution as well as for the intensity ratio have several physical origins and should be investigated in the future in more
detail. Finally, we want to remark that the derived results for the equatorial scan, namely the size distribution as well as the intensity distribution (log $\mu$, $\sigma$, contrast)
agree within errors with the results for the polar scan.

An interesting step in the future would be to do a quantitative comparison between the east-west with the south-north results. As no significant physical effects are
expected in east-west direction (at least for the quiet Sun) these data can be
used to calibrate the south-north scans. Such calibrated south-north data will allow similar investigations for small scale magnetic field activity, as those done for large scale magnetic fields (e.g sunspot cycle, activity belts).

\section*{Acknowledgements}
We are grateful to the \textit{Hinode} team for the possibility to use their data.
\textit{Hinode} is a Japanese mission developed and launched by ISAS/JAXA, with NAOJ as
domestic partner and NASA and STFC (UK) as international partners. It is
operated by these agencies in co-operation with ESA and NSC (Norway). The
research was funded by the Austrian Science Fund (FWF): J3176. D. U. and A. H.
are grateful to the {\"O}AD for financing a scientific stay at the Pic du Midi
Observatory. In addition D.U. and B.L. want to thank again the {\"O}AD and M\v{S}MT for financing a
short research stay at the Astronomical Institute of the Czech Academy of
Sciences in Ondrejov in the frame of the project MEB061109. Furthermore, J.J. wants to express his gratitude to the M\v{S}MT and {\"O}AD for financing a short research stay at the IGAM of the University of Graz.  M.R. is grateful to the Minist\`{e}re des Affaires
Etrang\`{e}res et Europ\'{e}ennes for financing a stay at the University of
Graz. D.U. is deeply grateful for the assistance given by J.C. del Toro Iniesta and H. Hudson while working on the manuscript. Finally, we want to acknowledge the support given by the anonymous
referee who helped us to improve this study.

\bibliographystyle{spr-mp-sola}
\bibliography{Literaturverzeichnis}

\end{article}

\end{document}